# Full-Duplex Aerial Communication System for Multiple UAVs with Directional Antennas

Tao Yu, Kiyomichi Araki, Kei Sakaguchi
Tokyo Institute of Technology, Japan
{yutao, araki, sakaguchi}@mobile.ee.titech.ac.jp

*Abstract*—UAV-based wireless systems, such as wireless relay and remote sensing, have attracted great attentions from academia and industry. To realize them, a high-performance wireless aerial communication system, which bridges UAVs and ground stations, is one of the key enablers. However, there are still issues hindering its development, such as the severe co-channel interference among UAVs, and the limited payload/battery-life of UAVs. To address the challenges, we propose an aerial communication system which enables system-level full-duplex communication of multiple UAVs with lower hardware complexities than ideal full-duplex communication systems. In the proposed system, each channel is re-assigned to the uplink and downlink of a pair of UAVs, and each UAV employ a pair of separated channels for its uplink and downlink. The co-channel interference between UAVs that reuse same channels is eliminated by exploiting advantages of UAVs' maneuverability and high-gain directional antennas equipped in UAVs and ground stations, so that dedicated cancellers are not necessary in the proposed system. The system design and performance analysis are given, and the simulation results well agree with the designs.

## I. Introduction

In the recent decade, unmanned aerial vehicles (UAVs) have evolved to be increasingly operable, functional, productive, and affordable than ever before thanks to the increasing miniaturization of electronic components and the development of high-performance flight control algorithms. They are attracting growing interests and popularity from fields of academia and industry because UAVs can flexibly extend their mobilities from 2D to 3D-space compared with conventional terrestrial robots. Such unique feature makes them into a suitable mobile platform for long-distance wireless systems requiring 3D-space mobilities, such as UAV-based wireless relay systems and UAV-based high-definition (HD) remote sensing systems, which are expected to be widely applied in e.g. infrastructure inspections, geodetic surveys, disasters relief, emergency communications, and monitoring systems [1]. The UAV and its derived applications are expected to have a global market of 127 billion USD by 2027 [2].

In order to take full advantages of UAV's 3D-space mobilities and evolve the UAV-based applications, real-time data transmission (e.g. HD sensing data, high-resolution video, and cellular traffic) via high-performance wireless aerial communication systems, which bridge UAVs and ground stations (GSs), is one of the key technologies and enablers. From the aspect of UAV-based remote systems, a desired aerial communication system requires features of large coverage, high throughput, and capability to simultaneously support multiple aerial links for multiple UAVs. On the other hand, from the aspect of UAV platform itself, due to limited battery-life and consequently limited payload weight/size, aerial communication modules with low hardware complexity and low energy consumption are desired.

However, there remains some challenges hindering the realization of aerial communications satisfying the abovementioned requirements. Due to the lack of natural or artificial obstacles such as hills and buildings in the working height of UAVs, line-of-sight (LOS) paths typically always exist in the wireless channels among UAVs. For multiple UAV systems, they lead to severe co-channel interference and the consequent extra system costs from such as guard interval, carrier sensing and collision avoidance, and make it inefficient to reuse the radio resource by the duplex and multiplex schemes such as time-division (TD) and frequency-division (FD). Therefore, spectrum efficiencies of the conventional multiple UAV aerial communication systems are always low, in contrast to the expanding required data rates of aerial communications and the limited frequency bands assigned to UAVs.

A wide range of research on UAV communication systems based on IEEE 802.11 families have been conducted by different researchers and summarized in [3]. The throughputs from 10 Mbps to 220 Mbps at distance from 50 m to 500 m were achieved. The performance of LTE network for the aerial communications has been studied by 3GPP in [4], and 150 Mbps throughput in experimental evaluation was reported [5]. Currently most of commercial aerial communication systems are employing Wi-Fi/LTE-based solutions or their derivatives. However, because both IEEE 802.11 families and LTE are originally designed for terrestrial usages, their system configurations, multiplex/duplex schemes, and interference avoidance mechanisms would be in low performance in the aerial communication systems. For example, the mechanism of carrier sensing multiple access with collision avoidance (CSMA/CA) would result in very low efficiency in the aerial propagation environments due to slow attenuation of interference and long propagation delays compared to terrestrial environments. Moreover, both Wi-Fi and LTE are semi-duplex systems with low spectrum efficiencies. MmWave-based aerial communication with ultra-large throughput and low co-channel interference is also an emerging field [6][7], but its coverage (typically up to hundreds of meters) is a natural defect for long-distance operations. Full-duplex systems can effectively increase the spectrum efficiency, and therefore the usages have been explored in some UAV-based applications, such as full-duplex UAV relays [8]-[11]. However, on the other hand, they suffer the severe self-interference and need extra dedicated hardware and energy for the interference cancellations [12][13], which are big burdens for UAVs. Moreover, all such systems generally are direct or simplified migrations of the terrestrial full-duplex communications without the special design for UAV-based systems.

To address the abovementioned challenges and meet the requirements for high-performance aerial wireless communications, we propose a full-duplex aerial communication system for multiple UAVs. A case study has been reported in [14] aiming to meet specific performance indicators from a practical surveillance system. In this paper, we generalize the system and show that it is applicable for UAV-based systems with varieties of communication requirements. In the proposed system, to avoid the co-channel interference among UAVs in the aerial propagation environments and increase the spectrum efficiency, highly directional antennas are employed in both GS and UAVs, and each frequency channel is re-used by the uplink and downlink of the pair of two different UAVs. Therefore, from the perspective of the system, all channels are simultaneously used for transmissions of uplinks and downlinks. Moreover, thanks to UAVs' capabilities of 3D mobilities, the hardware and processing of the co-channel interference canceller on UAVs is also avoided and replaced by actively controlling the relative positions of UAVs. Therefore, the high-performance full-duplex aerial communication with low hardware complexity and low co-channel interference is achieved by the spectrum separation and spatial multiplexing among UAVs.

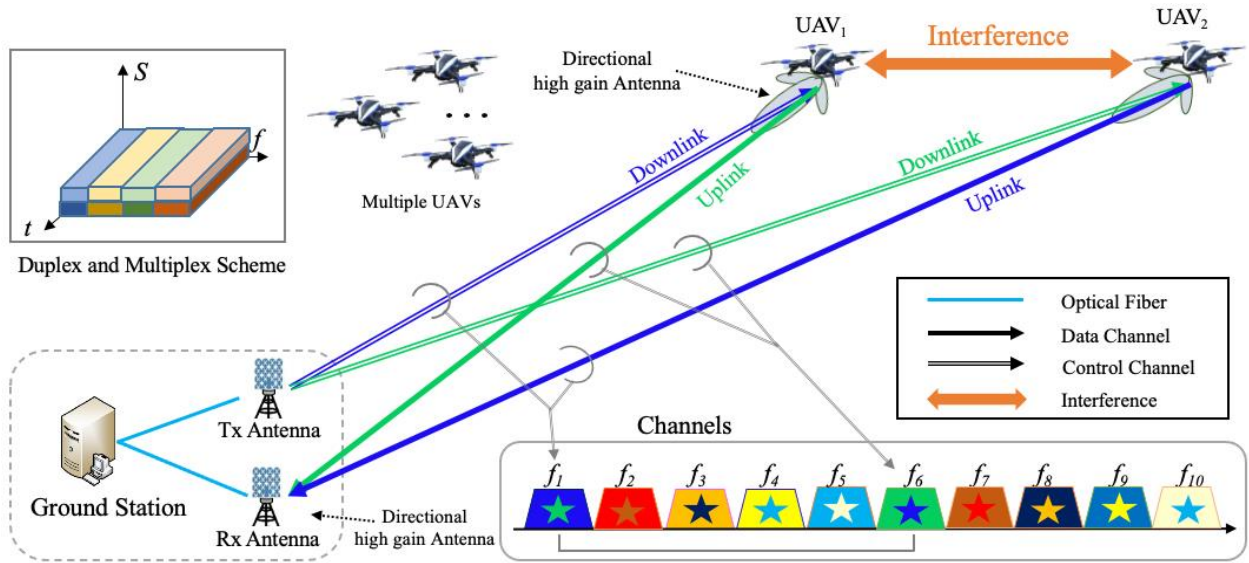

Fig.1. System architecture of the proposed aerial communication system for multiple UAVs.

This rest of this paper is organized as follows. Section II discusses the design methodologies and presents the architecture of the proposed system. Section IV analyzes the system performance by simulations. Finally, Section V concludes this paper.

## II. SYSTEM DESIGN

In this section, details of design methodology and system architecture of the proposed aerial communication system are given. The system aims to achieve high performance aerial communications with low hardware complexities.

### A. System Architecture

The architecture of the proposed full-duplex aerial communication system for multiple UAVs with directional antennas is illustrated in Fig. 1, in which technical details are explained by one pair of UAVs whose uplink and downlink use the same pair of channels. (In this paper, the uplink refers to a link from UAV to GS, and the downlink refers to the opposite.)

In each UAV in the proposed system, the uplink and downlink employ separated channels, so that complicated hardware, such as self-interference cancellers in the conventional full-duplex architecture for a single UAV, can be avoided. An example of the separated channels allocated for the uplink and downlink is also given in Fig.1, and the different colors represent the different frequency bands.

To increase the spectrum efficiency, each channel is reused by uplink of one UAV and downlink of another one. In Fig.1, the channel reallocation is illustrated by an example, in which the $UAV_1$ uses channel #1 (blue) and channel #6 (green) as the downlink and uplink respectively, while the $UAV_2$ uses them as the uplink and downlink, respectively. All other UAVs share two different channels in the same mechanism. Hence, all the channels are simultaneously reused by the uplinks and downlinks of the multiple UAVs in the proposed aerial communication system. The duplex and multiplex scheme is also illustrated in the upper left of Fig.1. The dark and light color represent the same channel reused by different UAVs, and the details will be explained in the next sub-section. Therefore, from the aspect of the whole system, a full-duplex aerial communication system is achieved.

One of the key issues to realize the system is to eliminate co-channel interference between a pair of UAVs which reuse the same channels, e.g. the interference from the uplinks of $UAV_1$ and $UAV_2$ to the downlinks of $UAV_2$ and $UAV_1$ in Fig. 1, which are especially severe in the aerial propagation environments. To address it, in the proposed system, highly directional antennas are equipped in both GS and UAVs. By the spatial multiplexing introduced by the highly directional antennas, the co-channel interference can be effectively eliminated, if the flight trajectories and relative positions of the pair of UAVs can be well controlled. Thanks to UAVs' capability of 3D-space mobilities, the hardware and processing cost of co-channel interference cancellations on UAVs are equated to a spatial control problem, which is particularly suitable for UAVs. The numerical analysis of the co-channel interference will be given in Sect. III.

### B. Duplex and Multiplex Schemes

To illustrate duplex and multiplex scheme in the proposed system, it is compared with the conventional schemes in Fig.2 which have widely used in the existing aerial systems to support the bi-directional multiple aerial links for multiple UAVs.

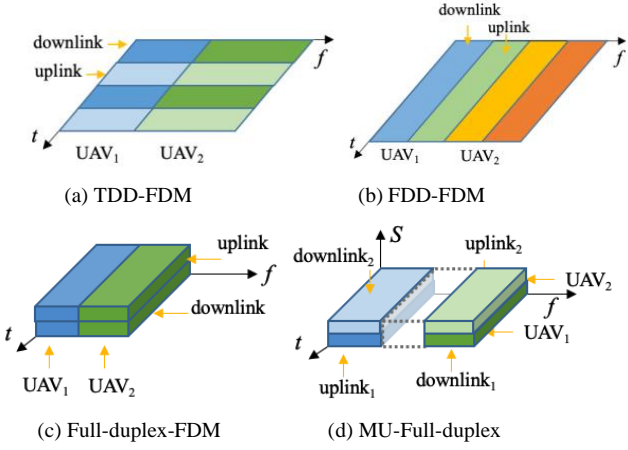

Fig.2. Duplex and multiplex schemes for aerial systems.

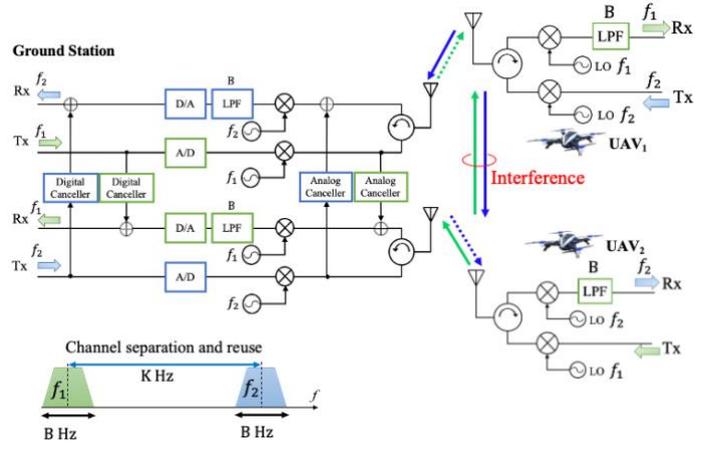

Fig.3. Low-complexity hardware architecture of the proposed system.

In the TD/FD duplex FD multiplex (TDD-FDM/FDD-FDM) systems as shown in Fig.2(a) and Fig.2(b), the interference can be well reduced because the non-overlapping time slots or channels are assigned to separate the up/downlinks and different UAVs, however, with the cost of low radio resource efficiency. Moreover, the delays brought by the TD schemes are especially harmful for the complex control systems such as UAVs. Fig.2(c) illustrates the full-duplex-FDM systems, which, ideally, have a high radio resource efficiency and well reduce the interference. To realize such systems, complicated hardware and signal processing are needed [12][13], which are big burdens for UAVs with highly limited battery-life and consequently limited payload. For example, high-performance analog-to-digital converters (ADCs) with high power consumption are necessary due to the significantly higher power of interference than signals of interests.

The proposed scheme aims to address the problems. The highly directional antennas introduce the spatial dimension as shown in Fig.2(d), which enables the spectrum sharing between the uplink and downlink of a pair of two different UAVs. The co-channel interference between them is obviously highly depends on their relative positions and antenna directions. Therefore, if the interference can be well eliminated by the active maneuvering of UAVs, the proposed system is expected to achieve a comparative performance with the ideal full-duplex-FDM system and has much higher radio resource efficiency than the TDD/FDD-FDM systems. Such mechanism is particularly suitable for aerial communication systems because it takes the full advantages of the unique capability of mobilities of UAVs.

### C. Beamforming

In the proposed system, beamforming must be performed due to the usage of highly directional antennas in both UAVs and GS. Despite the highly mobile nature of UAVs in 3D space, the beamforming and tracking is in fact still technically workable, because UAVs typically work in long-distance which results in a large beam coverage even for the highly directional antennas. For example, if GS employs a high gain directional antenna with a very narrow vertical beamwidth of 4°, the diameter of the vertical coverage would be 70 m and 350 m when the UAV is at 1 km and 5 km away from the GS, respectively. Therefore, the accuracies of e.g. GPS-based or high-zoom-camera-based localization and tracking of UAVs would be sufficient for location-based beamforming in the proposed system.

### D. Low-Complexity Hardware Architecture

Due to the complex hardware architecture of ideal full-duplex system, currently there is not any full-duplex UAV system on the market. The proposed system aims to achieve high-performance and high-efficiency aerial communication with low-complexity hardware. The architecture to realize the system is shown in Fig.3.

Comparing to GS, the hardware in the UAV side is quite simple. To achieve the lightweight and compact size, the transmitter and receiver share the common antenna. Because of the channel separation, the leakage from transmitter to the receiver can be well eliminated by low-pass filters. Regarding the co-channel interference, as illustrated between $UAV_1$ and $UAV_2$ in Fig.3, the proposed system removes the complicated hardware of co-channel cancellers by equating it to the active positions control of the UAV, while achieving a high radio resource efficiency.

It is noted that GS is employing complicated circuits and algorithms for interference cancelation by analog and digital canceller as shown in Fig.3, without size and energy limitations. We manly focus on the co-channel interference on UAVs in this paper.

## III. SYSTEM PERFORMANCE ANALYSIS

In this section, simulation configuration and performance analysis of the proposed system are given. For simplicity without loss of generality, only the performance of two UAVs which reuse the same channels are discussed in this section. However, the results also hold true for multiple UAVs, because the co-channel interference only occurs within these pairs.

The simulation model in this section for performance analysis is shown in Fig.4. Because LOS paths are assumed to exist between UAVs and GS, the two-ray model with ground reflection is used to simulate the air-to-ground and air-to-air channels, respectively. Therefore, the channel capacity of downlink of $UAV_2$, which is interfered by uplink of $UAV_1$, can be calculated as follows.

$$C = B \log_2 \left( 1 + \frac{P_{D2}^T \, P^L(\boldsymbol{p}_{Gs}, \boldsymbol{p}_2)}{N + P_{U1}^T \, P^L(\boldsymbol{p}_1, \boldsymbol{p}_2)} \right) \quad (1)$$

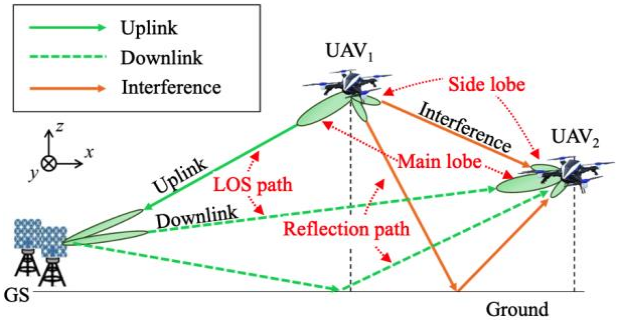

Fig.4. Simulation model for performance analysis.

where $B$ is the bandwidth of one channel, $N$ is noise, $P_{D2}^T$ is the transmission power of downlink from GS to $UAV_2$, $P_{U1}^T$ is the transmission power of uplink from $UAV_1$ to GS, $\boldsymbol{p}_{GS}, \boldsymbol{p}_1, \boldsymbol{p}_2$ are the positions of GS, $UAV_1$ and $UAV_2$. $P^L$ is the total gain modeled by two-ray model, which is a function of transmitter and receiver positions $\boldsymbol{p}_T$ and $\boldsymbol{p}_R$ as follows.

$$P^L(\boldsymbol{p}_T, \boldsymbol{p}_R) = \left[\frac{\lambda}{4\pi}\right]^2 \left|\frac{\sqrt{G^L(\boldsymbol{p}_T,\boldsymbol{p}_R)}}{D^L(\boldsymbol{p}_T,\boldsymbol{p}_R)} + \frac{R\sqrt{G^R(\boldsymbol{p}_T,\boldsymbol{p}_R)}e^{-j\Delta\phi_i}}{D^R(\boldsymbol{p}_T,\boldsymbol{p}_R)}\right|^2 \quad (2)$$

where $\lambda$ is wavelength, $D^L(\boldsymbol{p}_T, \boldsymbol{p}_R)$ and $D^R(\boldsymbol{p}_T, \boldsymbol{p}_R)$ is the lengths of LOS and reflection paths, $G^L(\boldsymbol{p}_T, \boldsymbol{p}_R)$ and $G^R(\boldsymbol{p}_T, \boldsymbol{p}_R)$ are the combined antenna gains of LOS and reflection paths, $\Delta\phi_i = 2\pi(D^R - D^L)/\lambda$ is phase difference between reflections and LOS path, $R$ is reflection coefficient.

Location-based beamforming is assumed to be performed so that the main lobes of the directional antennas on UAVs and GS are assumed to face each other with a random angle error $\sigma_{Beam} \sim N(0, 3°^2)$ which is used to model the beam misalignment in the calculation of combined antenna gains $G^L$ and $G^R$. The co-channel interference between UAVs is from the side lobes of the antennas. The antenna radiation pattern is modeled by *sinc*-function for analysis. The detailed simulation parameters are summarized in Table I.

### A. Analysis of the Achievable Capacity Performance

For convenience of illustration, channel capacities in the presence of the co-channel interference from $UAV_1$ to $UAV_2$ in the two types of positional relationship as shown in Fig.5 is analyzed. The two UAVs are in the same planes in y-axis and have a distance of $r$. The $UAV_2$ stays in a fix position of (5000, 0, 100). Namely, the horizontal distance between $UAV_2$ and GS is 5000 m, and its height is 100 m. In the linear positional relationship as shown in Fig.5(a), the two UAVs are in the same height, and in the circular positional relationship as shown in Fig.5(b), two UAVs are on the edge of a circle with the center at GS.

TABLE I  System parameters.

| Parameter | Value |
|---|---|
| Frequency | 5.7 GHz |
| Bandwidth | 10 MHz |
| Tx Power (GS/UAV) | 11 dBm / 0 dBm |
| GS antenna gain | 22 dBi |
| UAV antenna gain | 15 dBi |
| Antenna pattern | *sinc* function |
| BS beamwidth (ver./hor.) | 4° / 58° |
| UAV beamwidth (ver./hor.) | 36° / 36° |
| Air-to-ground channel | Two-ray |
| Air-to-air channel | Two-ray |
| Noise power density | -174 dBm/Hz |
| Noise factor | 5 dB |

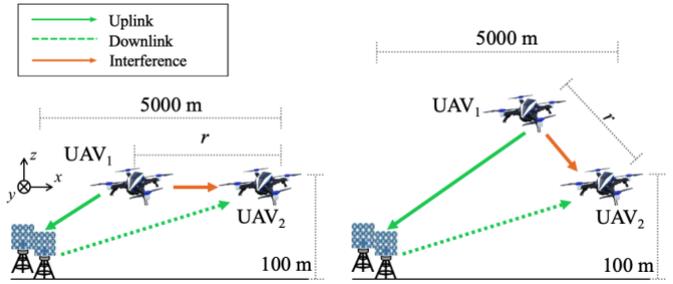

(a) Linear deployment  (b) Circular deployment

Fig.5. Positional relationships of UAVs for interference analysis.

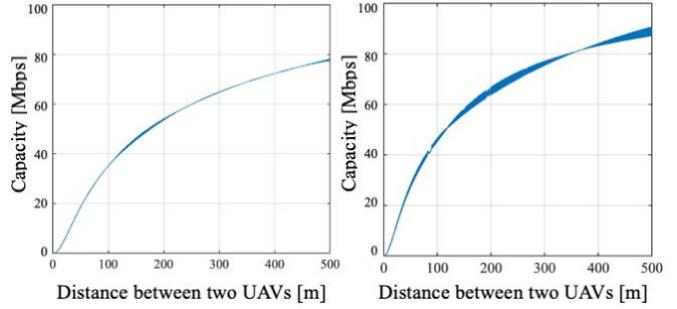

(a) Linear deployment  (b) Circular deployments

Fig.6. Downlink channel capacity vs distance between UAVs

The channel capacities of $UAV_2$ in the two positional relationships are given in Fig.6. The results show that the capacities highly depend on the distance and positional relationship between the two UAVs. Thanks to the proposed system architecture, very high capacities can be achieved if the two UAVs are separated by enough distance. For example, when the two UAVs are about 5000 m away from the GS as shown in Fig.5, if they are 50 m apart in linear or circular deployments, channel capacities of 20 Mbps or 28 Mbps can be achieved; if they 100 m apart in linear and circular deployments, channel capacities of 35 Mbps or 46 Mbps can be achieved. It shows the potentials of the proposed system in varieties of applications with different communication requirements. This can be achieved by different flight polices. The results also implies that high gain directional antennas effectively suppress the interference and increase the SINR. It also indicates that the high-performance aerial communication comparable to the ideal full-duplex system can be achieved by actively controller the positions of the two UAVs according to the required throughput.

### B. Analysis of System Efficiency

In this sub-section, the system efficiency including the power and radio resource efficiencies of the proposed system is compared with that of the widely used TDD-FDM systems.

The radio resource efficiency is defined as follows:

$$\eta = \frac{1}{R}\sum_{k=1}^{K}\left(R_U \log_2(1 + \frac{P_U^T P_k^L}{\sigma^2 + I_{U,k}}) + R_D \log_2\left(1 + \frac{P_D^T P_k^L}{\sigma^2 + I_{D,k}}\right)\right)$$

(3)

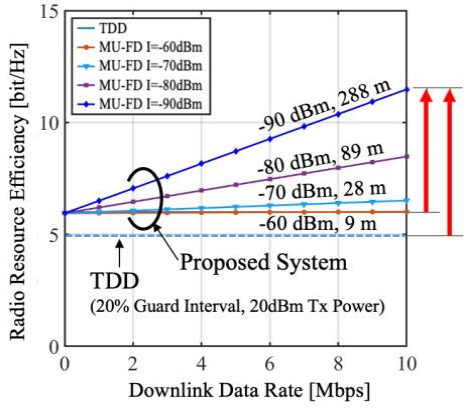

Fig.7. Radio resource efficiency.

where $R$, $R_U$, $R_D$ are the total, uplink and downlink radio resource, i.e. time slots or frequency bands, $k$ is the index of channel used by the UAV, $P_k^L$ is the total gain between GS and UAV using channel #$k$ including the antenna gains which obviously highly depends on the relative position of UAVs, $\sigma^2$ is the noise power, and $I_{u,k}$ and $I_{d,k}$ are the co-channel interference in uplink and downlink.

In the case of TDD-FDM system, $R = T = T_U + T_D + T_g$ is the frame length, $R_U = T_U$, $R_D = T_D$, and $T_U$, $T_D$, and $T_g$ are the length of time slots for uplink, downlink, and guard interval, respectively. Because each UAV uses a different channel, and orthogonal time slots are assigned for the uplink and downlink in each channel, it is assumed that there is no co-channel interference in both uplink and downlink, i.e. $I_{U,k} = 0$ and $I_{D,k} = 0$. In the simulation, the TDD-FDD system uses the omni-directional antennas with transmission power of 20 dBm. The guard interval $T_g$ is assumed to be 20% of the frame length $T$.

In the case of proposed multiple UAV full-duplex system, $R = B$ is total bandwidth of the system, and $R_U = R_D = B$ because all channels are reused by both uplinks and downlinks of UAVs. In the uplink, it is assumed that the GS can well cancel the co-channel interference by complicated hardware and algorithms, i.e. $I_{U,k} = 0$. However, in the downlink, because the interference from the uplink of the UAV using the same channel highly affects the system performance, the $I_{D,k}$ must be taken into the considerations in analysis. In the simulation, the parameters of the proposed system are the same as shown in Table I.

The radio resource efficiencies of the proposed system with different interference levels and the corresponding distances in linear deployment are given Fig.7. It shows that comparing to the conventional TDD-FDD systems, thanks to the channel reuse, the proposed system has a much higher radio resource efficiency, especially when the required downlink data rate is high. Moreover, it also shows that one of the key factors affecting the system performance is the co-channel interference level among UAVs. If it can be effectively eliminated, as show in the figure, the system radio resource efficiency can be increased.

In addition, the power efficiency is also an important factor for the energy critical platforms and systems such as UAVs. By introducing the high gain directional antennas on both of GS and UAVs, the power consumption for the uplink transmission from the UAV can be reduced. For example, with the assumption of the abovementioned parameters of the proposed system and TDD system with omni-directional antennas, in order to achieve the same data rate for uplink transmission, more than 20 dB transmission power can be saved on the UAV. The much lower transmission power does not only result in slower battery drain, but also smaller inter- and intra-system interference, which relax the flight areas of UAVs and the density of UAVs in one area.

*C. Interference Elimination by Flight Control*

As analyzed in the previous sections, due to the capability of 3D-space mobilities, the channel allocation and co-channel interference cancelation problems can be equated to the active flight trajectories and positions control problem in the proposed system. In this sub-section, the artificial potential method is used as an example of the UAV control algorithm for interference elimination. As in the previous sections, for simplicity, only two UAVs are controlled in the simulation, but the results also hold true for the case of multiple UAVs.

The artificial potential at a position is defined as the sum of the attractive, which is proportional to the distance to the destination, and repulsive force, which is modeled as the co-interference i.e. $I_{d,k}$.

$$U_n(t) = U_n^{\text{rep}}(\mathcal{L}_k(t)) + U_n^{\text{att}}(\boldsymbol{q}_n(t), \boldsymbol{q}_{Goal}) \quad (2)$$

in which,

$$U_n^{\text{att}}(\boldsymbol{q}_n(t), \boldsymbol{q}_{Goal}) = \omega D(\boldsymbol{q}_n(t), \boldsymbol{q}_{Goal})$$

$$U_n^{\text{rep}}(\mathcal{L}_k(t)) = I_{d,k}(\mathcal{L}_k(t)) = \sum_{\substack{j=1 \\ j \neq n}}^{J} P_{U,j}^T P^L(\boldsymbol{p}_j, \boldsymbol{p}_n)$$

where $\mathcal{L}_k(t) = \{\boldsymbol{q}_{k,1}(t), \boldsymbol{q}_{k,2}(t), \ldots\}$ is the set of positions of two UAVs using channel #$k$ in time $t$, and $\boldsymbol{q}_n(t)$ is the position of $\text{UAV}_n$ in time $t$, $\boldsymbol{q}_{Goal}$ is the position of the destination, $U_n^{\text{att}}$ is the attractive force to $\text{UAV}_n$, $\omega$ is the weight of attractive force, $D$ is the distance between two positions, $U_n^{\text{rep}}$ is the repulsive to $\text{UAV}_n$, $J$ is the number of UAVs reusing a channel (in this paper $J = 2$), $P_{U,j}^T$ is the uplink transmit power of $\text{UAV}_j$, $P^L(\boldsymbol{p}_j, \boldsymbol{p}_n)$ is the total gain between $\text{UAV}_j$ and $\text{UAV}_n$ including the path loss and gains of the directional antennas of UAVs.

Therefore, the flight trajectories of $\text{UAV}_n$ can be derived by minimizing the artificial potential field.

$$\hat{\boldsymbol{q}}_n(t+1) = \underset{\boldsymbol{q}_n(t+1) \in \mathcal{S}(t+1)}{\text{argmin}} U_n(t+1) \quad (3)$$

where $\mathcal{S}(t+1)$ is a set of positions defined by a simplified motion model of the UAV as follows.

$$\begin{aligned} -\boldsymbol{v}_{\max} &\leq \boldsymbol{v}(t+1) \leq \boldsymbol{v}_{\max} \\ -\boldsymbol{a}_{\max} &\leq \boldsymbol{a}(t+1) \leq \boldsymbol{a}_{\max} \end{aligned} \quad (4)$$

in which,

$$\boldsymbol{v}(t+1) = d\boldsymbol{q}_n(t+1)/dt \approx (\boldsymbol{q}_n(t+1) - \boldsymbol{q}_n(t))/\Delta t$$
$$\boldsymbol{a}(t+1) = d\boldsymbol{v}(t+1)/dt \approx (\boldsymbol{v}(t+1) - \boldsymbol{v}(t))/\Delta t$$

where $\boldsymbol{v}_{\max}$ and $\boldsymbol{a}_{\max}$ are decided by the hardware specifications of the UAV.

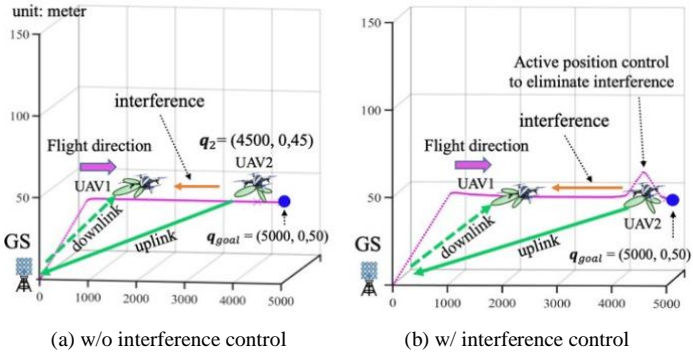

(a) w/o interference control  (b) w/ interference control

Fig.8. UAV flight trjactories.

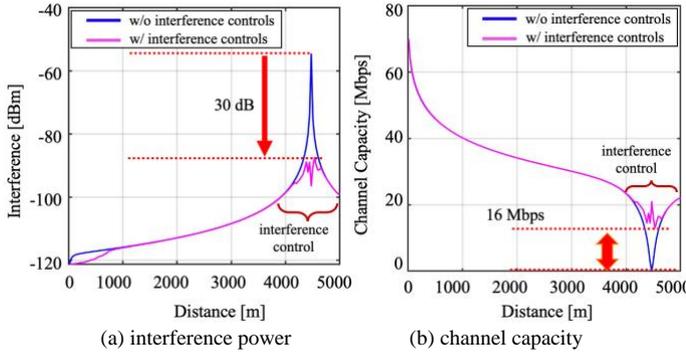

(a) interference power  (b) channel capacity

Fig.9 Performance comparasion of interference control.

For comparison, two simulations are conducted to evaluate whether the co-channel interference among UAVs can be eliminated. They are configured as shown in Fig.8(a) and Fig.8(b). In the simulations, $UAV_2$ hovers at position (4500, 0, 45), while $UAV_1$ flies to the destination (5000, 0, 50). The two UAVs reuse the same channels, so that the interference occurs from the uplinks to the downlinks. In the first simulation shown in Fig.8(a), no flight control is conducted, and $UAV_1$ directly flies to the destination. In the second simulation shown in Fig.8(b), active position control is performed on $UAV_1$ to eliminate the interference from the uplink of $UAV_2$.

Numerical results are presented in Fig.9, in which the pink and blue curves respectively represent the simulation scenarios with (w/) and without (w/o) interference controls, namely, flight trajectory control. In Fig.9(a) and Fig.9(b), the interference powers and the corresponding channel capacities are compared, respectively. The results clearly show that if the active flight control is performed when $UAV_1$ approaches $UAV_2$, co-channel interference as large as around 30 dB can be eliminated, and the channel capacity can be increased up to around 16Mbps. The performance improvement can be larger if the parameters $\omega$ in the control algorithm is adjusted to make the UAV perform maneuver in large scale, which should be configured according to the types of tasks and required date rates of UAVs.

## IV. CONCLUSION

In this paper, the authors proposed a design of a full-duplex aerial communication system for multiple UAVs with directional antennas. It aims at high radio resource efficiency and low hardware complexity on UAV by using the proposed channel reuse and interference elimination scheme. The architecture and design considerations are explained in detail. Numerical simulations are also conducted for validation. The simulation results confirmed higher performance and efficiency compared to the conventional TDD systems. The effectiveness of the interference elimination by flight control, instead of complicated hardware cancellers, is also verified by simulations.

There are still open issues in the system. Such research as the spectrum sharing schemes with other existing wireless systems and the more effective flight control algorithms for the resource allocation will be done in the future. In addition, the validation experiments are also under preparation, and are expected to be conducted soon.

ACKNOWLEDGMENT

The research leading to these results is funded by the Ministry of Internal affairs and Communications (MIC) of Japan under the grant agreements 0155-0083.